%% file: main.tex
\documentclass[twocolumn]{bmcart}% uncomment this for twocolumn layout and comment line below

\usepackage[hyphens]{url}  % DO NOT CHANGE THIS
\usepackage{graphicx} % DO NOT CHANGE THIS
\usepackage{caption}
\usepackage{algorithm}
\usepackage{algorithmic}

\usepackage{pifont}% http://ctan.org/pkg/pifont
\newcommand{\cmark}{\ding{51}}%
\newcommand{\xmark}{\ding{55}}%
\usepackage[hidelinks]{hyperref}
\usepackage{booktabs}       % professional-quality tables
\usepackage{amsfonts,amsmath,amssymb}
\usepackage{nicefrac}       % compact symbols for 1/2, etc.
\usepackage{xcolor}         % colors
\usepackage[nolist]{acronym}
\usepackage{ragged2e}

\DeclareMathOperator*{\argmax}{arg\,max}
\DeclareMathOperator*{\argmin}{arg\,min}

\usepackage{newfloat}
\usepackage{listings}
\DeclareCaptionStyle{ruled}{labelfont=normalfont,labelsep=colon,strut=off} % DO NOT CHANGE THIS
\floatstyle{ruled}
\newfloat{listing}{tb}{lst}{}
\floatname{listing}{Listing}
\begin{acronym}
    \input{sapacronyms.tex}
\end{acronym}

\usepackage[utf8]{inputenc} %unicode support
\graphicspath{ {./} }

\startlocaldefs
\endlocaldefs

\begin{document}

%%% Start of article front matter
\begin{frontmatter}

%\begin{fmbox}
\dochead{Methodology}

\title{Dual input neural networks for positional sound source localization}

\author[
  addressref={aff1},                   % id's of addresses, e.g. {aff1,aff2}
  %corref={aff1},                       % id of corresponding address, if any
  % noteref={n1},                        % id's of article notes, if any
  %email={e.grinstein@imperial.ac.uk}   % email address
]{\inits{E.G.}\fnm{Eric} \snm{Grinstein}}
\author[
  addressref={aff1},
%  email={vincent.neo09@imperial.ac.uk}
]{\inits{V.W.N.}\fnm{Vincent W.} \snm{Neo}}
\author[
  addressref={aff1},
 % email={p.naylor@imperial.ac.uk}
]{\inits{P.A.N.}\fnm{Patrick A.} \snm{Naylor}}

\address[id=aff1]{%                           % unique id
  \orgdiv{Department of Electrical and Electronic Engineering},             % department, if any
  \orgname{Imperial College London},          % university, etc
  \city{London},                              % city
  \cny{UK}                                    % country
}

\begin{abstractbox}

\begin{abstract} % abstract
\justifying
In many signal processing applications, metadata may be advantageously used in conjunction with a high dimensional signal to produce a desired output. In the case of classical Sound Source Localization (SSL) algorithms, information from a high dimensional, multichannel audio signals received by many distributed microphones is combined with information describing acoustic properties of the scene, such as the microphones' coordinates in space, to estimate the position of a sound source. We introduce Dual Input Neural Networks (DI-NNs) as a simple and effective way to model these two data types in a neural network. We train and evaluate our proposed DI-NN on scenarios of varying difficulty and realism and compare it against an alternative architecture, a classical Least-Squares (LS) method as well as a classical Convolutional Recurrent Neural Network (CRNN). Our results show that the DI-NN significantly outperforms the baselines, achieving a five times lower localization error than the LS method and two times lower than the CRNN in a test dataset of real recordings.
\end{abstract}

%%%%%%%%%%%%%%%%%%%%%%%%%%%%%%%%%%%%%%%%%%%%%%
%% Put each keyword in separate \kwd{}.     %%
%%%%%%%%%%%%%%%%%%%%%%%%%%%%%%%%%%%%%%%%%%%%%%


\end{abstractbox}

%\end{fmbox}% comment this for two column layout

\end{frontmatter}

\input{introduction}

\input{prior_art}

\input{method}

\input{experimentation}

\input{conclusion}

\section*{Appendix}

\input{appendix}
%%%%%%%%%%%%%%%%%%%%%%%%%%%%%%%%%%%%%%%%%%%%%%
%% Backmatter begins here                   %%
%%%%%%%%%%%%%%%%%%%%%%%%%%%%%%%%%%%%%%%%%%%%%%

\begin{backmatter}

% \section*{Acknowledgements}%% if any
% Not applicable.

\section*{Correspondence}
Eric Grinstein: e.grinstein@imperial.ac.uk

\section*{Funding}%% if any
This work was funded through the European Union's Horizon 2020 research and innovation programme under the Marie Skłodowska-Curie grant agreement no. 956369 and the UK Engineering and Physical Sciences Research Council (EPSRC) grant no. EP/S035842/1.

\section*{Abbreviations}%% if any
Dual Input Neural Network (DINN), convolutional recurrent neural network (CRNN), sound source localization (SSL)

% \section*{Availability of data and materials}%% if any
% The code repository \url{https://github.com/egrinstein/di_nn}, as well as a demonstration website \url{https://kaggle.com/code/egrinstein/di-nn-training-notebook} are made available as supplemental materials.

\section*{Ethics approval and consent to participate}%% if any
Not applicable.

% \section*{Competing interests}
% The authors declare that they have no competing interests.

% \section*{Consent for publication}%% if any
% Not applicable.

\section*{Authors' contributions}
Algorithmic development: Eric Grinstein, Patrick A. Naylor\\
Simulations and results: Eric Grinstein, Vincent W. Neo\\
Manuscript writing: Eric Grinstein, Vincent W. Neo, Patrick A. Naylor

% \section*{Authors' information}%% if any
% Text for this section\ldots

%%  After submission of the .TEX file,                     %%
%%  you will be prompted to submit your .BBL file.         %%

% if your bibliography is in bibtex format, use those commands:
\bibliographystyle{bmc-mathphys} % Style BST file (bmc-mathphys, vancouver, spbasic).
% \bibliography{bmc_article}      % Bibliography file (usually '*.bib' )
\bibliography{sapstrings,references}

\end{backmatter}
\end{document}

%% file: sapacronyms.tex
\acro{3GPP}{3rd Generation Partnership Project}
\acro{a-SLAM}[aSLAM]{Acoustic \aclu{SLAM}}
\acro{aSLAM}{Acoustic \aclu{SLAM}}
\acro{A-SNR}{A-weighted \acs{SNR}}
\acro{AAC}{Advanced Audio Coding\acroextra{. A lossy codec used for digital audio.}}
\acro{AAD}{Auditory Attention Detection}
\acro{AAS}{American Auditory Soc.}
\acro{AASP}{Audio and Acoustic Signal Processing}
\acro{ABC}{Analytical with or without Bias Compensation}
\acro{ABR}{Auditory-Brainstem Response}
\acro{ACAWD}{Archivable Core Actual-Word Database}
\acro{ACB}{Adaptive Codebook}
\acro{ACC}{Accuracy}
\acro{ACE}{Acoustic Characterization of Environments\acroextra{. A noisy reverberant speech corpus and IEEE challenge run by the SAP group at Imperial College}}
\acro{ACELP}{Algebraic Code-Excited Linear Prediction}
\acro{ACF}{Autocorrelation Function}
\acro{ACK}{Acknowledgement}
\acro{ACL}{Access Control List}
\acro{ACR}{Absolute Category Rating}
\acro{AD}{Audio Diarization}
\acro{ADC}{Analogue-to-Digital Converter}
\acro{ADM}{Adaptive Differential Microphone}
\acro{ADPCM}{Adaptive Differential Pulse Code Modulation}
\acro{ADSL}{Asymmetric Digital Subscriber Line}
\acro{AE}{Almost Everywhere}
\acro{AES}{Audio Engineering Society}
\acro{AES2}[AES]{Advanced Encryption Standard}
\acro{AGC}{Automatic Gain Control}
\acro{AH}{Amplitude Histogram}
\acro{AI}{Articulation Index}
\acro{AI2}[AI]{Artificial Intelligence}
\acro{AI3}[AI]{Audio Inpainting}
\acro{AIC}{Akaike Information Criterion}
\acro{AIFF}{Audio Interchange File Format}
\acro{AIR}{Acoustic Impulse Response}
\acro{AIR2}[AIR]{Aachen Impulse Response}
\acro{AIRD}{Aachen Impulse Response Database}
\acro{ALC}{Automatic Level Control}
\acro{ALCons}{Articulation Loss of Consonants}
\acro{AM}{Amplitude Modulation}
\acro{AMDF}{Average Magnitude Difference Function\acroextra{. A function with similar properties to the cross- or autocorrelation but that requires no multiplication to evaluate.}}
\acro{AMR}{Adaptive Multi-Rate}
\acro{AMR-NB}{Adaptive Multi-Rate Narrow Band}
\acro{AMR-WB}{Adaptive Multi-Rate Wide Band}
\acro{AMS}{Amplitude Modulation Spectrogram}
\acro{ANC}{Adaptive Noise Canceller}
\acro{ANS}{Autocorrelation-Based Noise Subtraction}
\acro{ANSI}{American National Standards Institute}
\acro{ANU}{Australian National University}
\acro{APLAWD}{Archivable Priority List Actual-Word Database}
\acro{AR}{Autoregressive}
\acro{ARD}{Arbeitsgemeinschaft der \"{o}ffentlich-rechtlichen Rundfunkanstalten der Bundesrepublik Deutschland\acroextra{. `Consortium (``Working group'') of the public-law broadcasting institutions of the Federal Republic of Germany'}}
\acro{ARMA}{Autoregressive Moving Average}
\acro{AS}{Audio Segmentation}
\acro{AS2}[AS]{Almost Surely}
\acro{ASA}{Acoustic Scene Analysis}
\acro{ASIO}{Audio Stream Input/Output\acroextra{. A computer soundcard protocol with low latency developed by Steinberg.}}
\acro{ASK}{Amplitude Shift Keying}
\acro{ASLP}{Audio, Speech, and Language Processing}
\acro{ASM}{Acoustic Scene Mapping}
\acro{ASR}{Automatic Speech Recognition}
\acro{ASS}{Approximate Spectrum Substitution}
\acro{AST}{Acoustic Source Tracking}
\acro{AST2}[AST]{Asymmetric Sampling in Time}
\acro{ATF}{Acoustic Transfer Function\acroextra{. The Fourier Transform of the \acs{RIR}.}}
\acro{ATLM}{Acoustic Tokenization and Language Modelling}
\acro{ATR}{Advanced Telecommunications Research Institute International\acroextra{, Kyoto, Japan}}
\acro{AUC}{Area under the curve}
\acro{AURORA}{Aurora Experimental Framework for the Evaluation of the Performance of Speech Recognition Systems under Noisy Conditions}
\acro{AUV}{Autonomous Underwater Vehicle}
\acro{AV}{Audio-Visual}
\acro{AWGN}{Additive White Gaussian Noise}
\acro{AWS}{Approximate Waveform Substitution}
\acro{BASIE}{Bayesian Adaptive Speech Intelligibility Estimation}
\acro{BCE}{Blind Channel Estimation}
\acro{BCL}{Bekesy Comfortable Loudness}
\acro{BCR}{Block-Coordinate Relaxation}
\acro{BEM}{Boundary Element Method}
\acro{BER}{Bit Error Rate}
\acro{BIBO}{Bounded-Input Bounded-Output}
\acro{BIC}{Bayesian Information Criterion}
\acro{Bk}{Berksons\acroextra{. A unit for measuring intelligibility.}}
\acro{BK}[B\&K]{Br{\"u}el and Kj{\ae}r}
\acro{BLSTM}{Bidirectional \acs{LSTM}}
\acro{BM}{Blocking Matrix}
\acro{BO-SCPHD}{Bearing-only \acs{SC-PHD}}
\acro{BO-SLAM}{Bearing-only \acs{SLAM}}
\acro{BOT}{Bearing-only tracking}
\acro{BP}{Basis Pursuit}
\acro{BPCC}{Basis Pursuit with Clipping Constraints}
\acro{BPDN}{Basis Pursuit Denoising}
\acro{BPM}{Beats Per Minute}
\acro{BPSK}{Binary Phase Shift Keying}
\acro{BR}{Barrodale and Roberts' (algorithm)}
\acro{BRI}{Basic Rate Index}
\acro{BSD}{Bark Spectral Distortion}
\acro{BSI}{Blind System Identification}
\acro{BSIM}{Binaural Speech Intelligbility Model}
\acro{BSS}{Blind Source Separation}
\acro{BSTOI}{Binaural \acs{STOI}}
\acro{BW}{Bandwidth}
\acro{BZ}{Back-to-Zero}
\acro{C4DM}{Centre for Digital Music}
\acro{C50}[$C_\textrm{50}$]{Clarity Index}
\acro{C-GFB}{Combination Gas-fired Boiler}
\acro{CART}{Classification and Regression Tree}
\acro{CASA}{Computational Auditory Scene Analysis}
\acro{CBR}{Constant Bit Rate}
\acro{CCC}{Cross-Correlation Coefficient}
\acro{CCCC}{DARPA CSR Corpus Coordinating Committee}
\acro{CCD}{Charge-Coupled Device}
\acro{CCI}{Call Clarity Index}
\acro{CCITT}{Consultative Committee for International Telephony and Telegraphy}
\acro{CCM}{Contralateral Competing Message}
\acro{CCR}{Comparison Category Rating}
\acro{CDB}{Constant Directivity Beamformer}
\acro{CDMA}{Code Division Multiple Access}
\acro{CELP}{Code-excited Linear Prediction}
\acro{CHIEF}{Combined Helmholtz Integral Equation Formulation}
\acro{CIT}{Constrained Initial Taps}
\acro{CL}{Clipping Level}
\acro{CLEAR}{Centre for Law Enforcement Audio Research}
\acro{CLID}{Cluster Identification Test}
\acro{CLT}{Central Limit Theorem}
\acro{CMA}{Constant Modulus Algorithm}
\acro{CMASI}{Coherence Modulated Acoustic Speckle Interferometry}
\acro{CMB}{Cosmic Microwave Background}
\acro{CNC}{Consonant-Nucleus-Consonant}
\acro{CNG}{Comfort Noise Generation}
\acro{CNN}{Convolutional Neural Network}
\acrodefplural{CNN}[CNNs]{Convolutional Neural Networks}
\acro{CODEC}{Coder-Decoder}
\acro{CPE}{Customer Premises Equipment}
\acro{CPHD}{Cardinalized \acs{PHD}}
\acro{CRACD}{Codec-robust Automatic Clipping Detector}
\acro{CRC}{Cyclic Redundancy Check}
\acro{CRNN}{Convolutional Recurrent Neural Network}
\acrodefplural{CRNN}[CRNNs]{Convolutional Recurrent Neural Networks}
\acro{CS}{Channel Shortening}
\acro{CS2}[CS]{Compressive Sensing}
\acro{CSP}{Communications and Signal Processing}
\acro{CSR-WSJ}{Continuous Speech Recognition Wall Street Journal Phase 1\acroextra{ database}}
\acro{CST}{Connected Speech Test\acroextra{ speech corpus}}
\acro{CT}{Conversation Test}
\acro{CTTN}{Comparative Tolerance to Noise}
\acro{CV}{Coefficient of Variation}
\acro{CVNN}{Complex-Valued Neural Networks}
\acrodefplural{CV}{Coefficients of Variation}
\acro{CV2}[CV]{Constant Velocity}
\acro{CVC}{Consonant-Vowel-Consonant}
\acro{CW}{Continuous Wave}
\acro{CWM}{Centre-Weighted Median}
\acro{CWMY}{Centre-Weighted Myriad}
\acro{CWT}{Continuous Wavelet Transform}
\acro{DAC}{Digital-to-Analogue Converter}
\acro{DAM}{Diagnostic Acceptability Measure}
\acro{DAQ}{Data Acquisition}
\acro{DARPA}{Defense Advanced Research Projects Agency\acroextra{ of the United States Dept. of Defense}}
\acro{DARPA-RMD}{\acs{DARPA} 1000-Word Resource Management Database\acroextra{ for Continuous Speech Recognition}}
\acro{DAW}{Digital Audio Workstation}
\acro{dB}{Decibel}
\acro{dBFS}{\acs{dB} Full Scale}
\acro{DBN}{Deep Belief Network}
\acro{DBSTOI}{Deterministic \acs{BSTOI}}
\acro{DC}{Direct Current}
\acro{DCME}{Digital Circuit Multiplexing Equipment}
\acro{DCR}{Degradation Category Rating}
\acro{DCT}{Discrete Cosine Transform}
\acro{DDR}{Direct-to-diffuse ratio}
\acro{DDR3}{Double Data Rate Type Three}
\acro{DECT}{Digital European Cordless Telecommunication}
\acro{DeLILAH}{Detection of Clipping using Least Squares Residuals and Iterated Logarithm Amplitude Histogram}
\acro{DENBE}{\acs{DRR} Estimation using a Null-Steered Beamformer}
\acro{DET}{Detection Error Trade-off}
\acro{Dev}{Development\acroextra{ dataset of the \acs{ACE} Challenge}}
\acro{DFT}{Discrete Fourier Transform}
\acro{DI}{Directivity Index}
\acro{DI-NN}{Dual-Input Neural Network}
\acro{DirectX}{\acroextra{A programming interface developed by Microsoft for handling tasks related to multimedia.}}
\acro{DIRHA}{Distant-speech Interaction for Robust Home Applications\acroextra{ multi-microphone multi-language acoustic speech corpus}}
\acro{DMA}{Differential Microphone Array}
\acro{DMT}{Discrete Multi-Tone}
\acro{DMV}{Dynamically Managed Voice\acroextra{ system}}
\acro{DNN}{Deep Neural Network}
\acro{DNR}{Dynamic Noise Reduction}
\acro{DoA}{Direction-of-Arrival}
\acrodefplural{DoA}[DoAs]{Directions-of-Arrival}
\acro{DOA}{Direction-of-Arrival}
\acrodefplural{DOA}[DOAs]{Directions-of-Arrival}
\acro{DP}{Dynamic Programming}
\acro{DPCM}{Differential Pulse Code Modulation}
\acro{DPD}{Direct-Path Dominance}
\acro{DPD-MUSIC}{Direct-Path Dominance Multiple Signal Classification}
\acro{DR}{Douglas-Rachford}
\acro{DR2}{Dynamic Range}
\acro{DRM}{Diagnostic Rhyme Test}
\acro{DRR}{Direct-to-Reverberant Ratio}
\acro{DRNN}{Deep Recurrent Neural Net}
\acro{DRT}{Diagnostic Rhyme Test}
\acro{DSB}{Delay-and-Sum Beamformer}
\acro{DSOBM}{Deterministic \acs{SOBM}}
\acro{DSP}{Digital Signal Processing}
\acro{DSPS}{Double Sides Periodic Substitution}
\acro{DSR}{Distributed Speech Recognition}
\acro{DSWS}{Double Sides Waveform Substitution}
\acro{DTMF}{Dual Tone Multi-Frequency}
\acro{DTX}{Discontinued Transmission}
\acro{DWT}{Discrete Wavelet Transform}
\acro{EARS}{Embodied Audition for RobotS}
\acro{EBF}{Eigen-beamformer}
\acro{EBU}{European Broadcasting Union}
\acro{EC}{Echo Canceller}
\acro{EDC}{Energy Decay Curve}
\acro{EDR}{Energy Decay Relief}
\acro{EDF}{Energy Decay Function}
\acro{EEG}{Electroencephalography}
\acro{EER}{Equal Error Rate}
\acro{EFICA}{Efficient Fast Independent Component Analysis}
\acro{EIR}{Equalized Impulse Response}
\acro{EKF}{Extended Kalman Filter}
\acro{EKF-SLAM}[EKF-SLAM]{\acs{EKF} \acs{SLAM}}
\acro{EKF-SLAM2}[EKF-SLAM]{Extended Kalman Filter \acs{SLAM}}
\acro{EL}{Echo Loss}
\acro{ELF}{Extremely Low Frequency}
\acro{ELRA}{European Languages Research Association}
\acro{EM}{Estimation-Maximization\acroextra{. An iterative technique to solve certain optimization problems.}}
\acro{EMIB}{Eigenmike Microphone Interface Box}
\acro{ENF}{Electrical Network Frequency}
\acro{EPSRC}{Engineering and Physical Sciences Research Council}
\acro{EQ}{Equalisation}
\acro{ERB}{Equivalent Rectangular Bandwidth}
\acro{ERP}{Ear Reference Point (cf. ITU-T Rec. P.64 1999)}
\acro{ESA}{Early Stage Assessment}
\acro{ESM}{Equivalent Source Method}
\acro{ESPRIT}{Estimation of Signal Parameters via Rotational Invariance Techniques}
\acro{ETAN}{Equivalent Tolerance to Additional Noise} 
\acro{ETSI}{European Telecommunications Standards Institute}
\acro{EURASIP}{European Association for Signal Processing}
\acro{EUSIPCO}{European Signal Processing Conference}
\acro{Eval}{Evaluation\acroextra{ dataset of the \acs{ACE} Challenge}}
\acro{F1}{F1 Score}
\acro{FAR}{False Alarm Rate}
\acro{FastSLAM}[FastSLAM]{FActored Solution To Simultaneous Localization and Mapping}
\acro{FastSLAM2}[FastSLAM]{FActored Solution To \acs{SLAM}} 
\acro{FAU}{Friedrich-Alexander-Universit{\"a}t}
\acro{FB}{Forward-Backward}
\acro{FB2}[FB]{Fullband}
\acro{FBF}{Fixed Beamformer}
\acro{FBSS}{Forward-Backward Spatial Smoothing}
\acro{FCC}{Federal Communications Commission}
\acro{FC-NN}{Fully Connected Neural Network}
\acro{FDM}{Frequency Division Multiplexing}
\acro{FDR}{False Discovery Rate}
\acro{FDR2}[FDR]{Free-Decay Region}
\acro{FEC}{Forward Error Correction}
\acro{FEM}{Finite Element Method}
\acro{FFI}{Norwegian Defence Research Establishment}
\acro{FFT}{Fast Fourier Transform}
\acroindefinite{FFT}{an}{a}
\acro{FIFO}{First-In First-Out}
\acro{FIR}{Finite Impulse Response\acroextra{. A filter whose output is a weighted sum of past input values and whose system function contains only zeros and no poles.}}
\acro{FISM}{Fast Image Source Method}
\acro{FISST}{Finite Set STatistics}
\acro{FLOM}{Fractional Lower-Order Moments}
\acro{FLOS}{Fractional Lower-Order Statistics}
\acro{FM}{Frequency Modulation}
\acro{FMM}{Fast Multipole Method}
\acro{FN}{False Negative}
\acro{FN2}{Nth Formant}
\acro{FNR}{False Negative Rate}
\acro{FORTRAN}{The IBM Mathematical Formula Translating System}
\acro{FoV}{Field of View}
\acro{FP}{False Positive}
\acro{FPR}{False Positive Rate}
\acro{FPS}{Frames Per Second}
\acro{FRI}{Finite Rate of Innovation}
\acro{FSB}{Filter-and-Sum Beamformer}
\acro{FSK}{Frequency Shift Keying}
\acro{FT}{Flat-Top}
\acro{FWER}{Familywise Error Rate}
\acro{FWSSNR}{Frequency-Weighted Segmental \acs{SNR}}
\acro{FWSSRR}{Frequency-Weighted \acs{SSRR}}
\acro{G.711}{\acs{PCM} of Voice Frequencies}
\acro{GARCH}{Generalized Auto-regressive Conditional Heteroscedasticity}
\acro{GBW}{Gain Bandwidth Product}
\acro{GCC}{Generalized Cross-Correlation}
\acro{GCC-PHAT}{Generalized Cross-Correlation with Phase Transform\acroextra{ method of estimating \acs{TDoA}}}
\acro{GCF}{Global Coherence Field}
\acro{GGD}{Generalized Gaussian Distribution}
\acro{GGD2}[G$\Gamma$D]{Generalised Gamma Distribution}
\acro{GM}{Gaussian Mixture}
\acro{GM-PHD}{Gaussian Mixture \acs{PHD}}
\acro{GMCA}{Generalized Morphological Component Analysis}
\acro{GMM}{Gaussian Mixture Model\acroextra{. An approximation to an arbitrary probability density function that consists of a weighted sum of Gaussian distributions}}
\acro{GNSS}{Global Navigation Satellite System}
\acro{GPRS}{General Packet Radio Services}
\acro{GPS}{Global Positioning System}
\acro{GSC}{Generalized Sidelobe Canceller}
\acro{GSM}{Global System for Mobile Communications}
\acro{GSM-EFR}{\acs{GSM} Enhanced Full Rate Codec}
\acro{GSM-FR}{\acs{GSM} Full Rate Codec}
\acro{GSM-HR}{\acs{GSM} Half Rate Codec}
\acro{GUI}{Graphical User Interface}
\acro{HAAC}{High Amplitude Audio Capture}
\acro{HATS}{Head and Torso Simulator}
\acro{HERB}{Harmonicity-based dEReverBeration}
\acro{HFT}{Hands-Free Terminal}
\acro{HI}{Hearing-Impaired}
\acro{HIE}{Helmholtz Integral Equation}
\acro{HISAS}{High resolution Interferometric \ac{SAS}}
\acro{HINT}{Hearing-in-Noise Test}
\acro{HLT}{Human Language Technology}
\acro{HMM}{Hidden Markov Model}
\acro{HOS}{Higher-Order Statistics}
\acro{HPF}{High-Pass Filter}
\acro{HR}{Half Rate (\acs{GSM} Codec)}
\acro{HRI}{Human-Robot Interaction}
\acro{HRTF}{Head-related Transfer Function}
\acro{HSA}{Hearing, Speech, Audio\acroextra{ technology group at Fraunhofer IDMT}}
\acro{HSD}{Hybrid Steepest Descent}
\acro{HT}{Hannan-Thomson}
\acro{HTK}{Hidden Markov Model Tool Kit}
\acro{IBM}{Ideal Binary Mask}
\acro{IC}{Interference Canceller}
\acro{ICA}{Independent Component Analysis}
\acro{ICASSP}{Intl. Conf. on Acoustics, Speech and Signal Processing}
\acro{ID}{Identifier}
\acro{IDMT}{Institute for Digital Media Technology}
\acro{iDEN}{Integrated Digital Enhanced Network}
\acro{IEC}{International Electrotechnical Commission}
\acro{IEEE}{Institute of Electrical and Electronics Engineers}
\acro{IET}{Institute of Engineering and Technology}
\acro{IETF}{Internet Engineering Task Force}
\acro{IFFT}{Inverse Fast Fourier Transform}
\acro{IHC}{Inner Hair Cell}
\acro{IHT}{Iterative Hard Thresholding}
\acro{IID}{Independent and Identically Distributed}
\acro{IIR}{Infinite Impulse Response\acroextra{. A filter whose output is a weighted sum of both past input and past output values and whose system function contains both poles and zeros.}}
\acro{IL}{Iterated Logarithm\acroextra{, the logarithm of the logarithm}}
\acro{ILAH}{Iterated Logarithm Amplitude Histogram\acroextra{ clipping detection method}}
\acro{ILD}{Interaural Level Difference}
\acro{IMCRA}{Improved Minima Controlled Recursive Averaging\acroextra{. A technique for blindly estimating the spectrum of additive noise in a signal.}}
\acro{IMD}{Inter-Modulation Distortion}
\acro{IMSI}{International Mobile Subscriber Identity}
\acro{IMU}{Inertial Measurement Unit}
\acro{INMD}{In-service Non-intrusive Measurement Device}
\acro{INTERSPEECH}{Annual Conference of the \acs{ISCA}}
\acro{IO}{Infinitely Often}
\acro{IP}{Internet Protocol}
\acro{IPA}{International Phonetic Association}
\acro{IPD}{Interaural Phase Difference}
\acro{IRM}{Ideal Ratio Mask}
\acro{IRS}{Inverse repeated Sequence\acroextra{. A pseudo random sequence used for impulse response measurement.}}
\acro{IRS2}[IRS]{Intermediate Reference System}
\acro{IS}{Importance Sampling}
\acro{ISCA}{International Speech Communication Association}
\acro{ISDN}{Integrated Services Digital Network}
\acro{ISFT}{Inverse \acl{SFT}}
\acro{ISO}{Intl. Organization for Standardization}
\acro{ISP}{Intensity Spectral Profile}
\acro{IST}{Iterative Soft Thresholding}
\acro{ISTFT}{Inverse Short Time Fourier Transform}
\acro{ISVR}{Institute of Sound and Vibration Research\acroextra{, Southampton University, UK}}
\acro{ITD}{Interaural Time Difference}
\acro{ITF}{Interaural Transfer Function}
\acro{ITU}{International Telecommunication Union}
\acro{ITUR}[ITU-R]{International Telecommunication Union\acroextra{ Radiocommunication Sector}}
\acro{ITUT}[ITU-T]{International Telecommunication Union\acroextra{ Telecommunication Standardisation Sector}}
\acro{IUWT}{Isotropic Undecimated Wavelet (Starlet) Transform}
\acro{IWAENC}{Intl. Workshop on Acoustic Signal Enhancement}
\acro{IWAENC_PRE_2012}[IWAENC]{Intl. Workshop Acoustic Echo and Noise Control}
\acro{IWASE}{Intl. Workshop on Acoustic Signal Enhancement}
\acro{JADE}{Joint Approximate Diagonalization of Eigen-Matrices}
\acro{JPDA}{Joint Probabilistic Data Association}
\acro{JPEG}{Joint Photographic Experts Group}
\acro{KDE}{Kernel Density Estimate}
\acro{KF}{Kalman Filter}
\acro{KFD}{Kernel Fisher Discriminant\acroextra{ Analysis}}
\acro{KL}{Karhunen-Lo{\'{e}}ve}
\acro{KLT}{Karhunen-Lo{\'{e}}ve Transform}
\acro{KST}{Kolmogorov-Smirnov Test}
\acro{LAD}{Least Absolute Deviation}
\acro{LAN}{Local Area Network}
\acro{LARS}{Least Angle Regression}
\acro{LAT}[L$_{\textrm{AT}}$]{Equivalent Continuous Sound Level\acroextra{. Also called Leq}}
\acro{LBR}{Low Bitrate Redundancy}
\acro{LC}{Local Criterion}
\acro{LCMP}{Linearly Constrained Minimum Power}
\acro{LCMV}{Linearly Constrained Minimum Variance}
\acro{LCWM}{Linear Combination of Weighted Medians}
\acro{LDC}{Linguistic Data Consortium}
\acro{LEM}{Loudspeaker-Enclosure-Microphone System}
\acro{Leq}[L$_{\textrm{eq}}$]{Equivalent Continuous Sound Level\acroextra{. Also called LAT}}
\acro{LF}{Liljencrants-Fant\acroextra{. The developers of a glottal waveform model}}
\acro{LHS}{Left-Hand Side}
\acro{LID}{Language Identification}
\acro{LILAH}{\acs{LSR2}-\acs{ILAH}\acroextra{ clipping detection method}}
\acro{LIME}{LInear Predictive Multi-input Equalization\acroextra{ algorithm}}
\acro{LiNoPS}{Lightweight Noise Protection System}
\acro{LLN}{Law of Large Numbers}
\acro{LLR}{Log-Likelihood Ratio}
\acro{LLS}{Logarithmic Least Squares}
\acro{LMA}{Least Mean Absolute}
\acro{LMS}{Least Mean Squares\acroextra{ adaptive filter}}
\acro{LNA}{Low Noise Amplifier}
\acro{LoS}{Line-of-Sight}
\acro{LOT}{Listening-Only Test}
\acro{LP}{Linear Parameter}
\acro{LP2}[LP]{Linear Predictive}
\acro{LP3}[LP]{Linear Prediction}
\acro{LPC}{Linear Predictive Coding\acroextra{. An autoregressive model of speech production.}}
\acro{LQO}{Listening Quality Objective}
\acro{LREC}{Conf. on Language Resources and Evaluation}
\acro{LS}{Least-Squares}
\acro{LSA}{Log Spectral Amplitude}
\acro{LSB}{Lower Side-Band}
\acro{LSB2}[LSB]{Least Significant Bit}
\acro{LSD}{Log Spectral Distortion}
\acro{LSP}{Line Spectrum Pairs}
\acro{LSR}{Late Stage Review}
\acro{LSR2}[LSR]{Least Squares Residuals\acroextra{ clipping detection method}}
\acro{LSRT}{Least Squares Residuals with Thresholding\acroextra{ clipping detection method}}
\acro{LSTM}{Long Short-Term Memory}
\acro{LTASS}{Long Term Average Speech Spectrum}
\acro{LTI}{Linear Time Invariant}
\acro{LTP}{Long Term Prediction}
\acro{LU}{Loudness Unit}
\acro{LUFS}{Loudness Units Full-Scale}
\acro{MA}{Moving Average}
\acro{MAC}{Multiply Accumulate Operation}
\acro{MAD}{Median Absolute Deviation}
\acro{MAE}{Mean Absolute Error}
\acro{MAP}{Maximum \emph{a posteriori}}
\acro{MARDY}{Multichannel Acoustic Reverberation Database at York}
\acro{MARS}{Multivariate Adaptive Regression Splines}
\acro{MBF}{Matched Filter Beamformer}
\acro{MC}{Monte Carlo}
\acro{MCA}{Morphological Component Analysis}
\acro{MCC}{Matthew's Correlation Coefficient}
\acro{MCEQ}{MultiChannel EQualisation}
\acro{MCS}{Multidimensional Colouration Space}
\acro{MDCT}{Modified Discrete Cosine Transform}
\acro{MDL}{Minimum Description Length}
\acro{MDS}{Multidimensional Scaling}
\acro{Mel}{\acroextra{A non-uniform frequency scale corresponding to perceived frequency. It is approximately linear at low frequencies and logarithmic at high frequencies.}}
\acro{MELP}{Mixed Excitation Linear Prediction}
\acro{MFCC}{Mel-frequency Cepstral Coefficients}
\acro{MFS}{Method of Fundamental Solutions}
\acro{MFSK}{Multi-Frequency Shift Keying}
\acro{MHT}{Multi-Hypotheses Tracking}
\acro{MI}{Mutual Information}
\acro{MIMO}{Multiple-Input-Multiple-Output}
\acro{MINT}{Multiple-input/output INverse Theorem}
\acro{MIRS}{Motorola Integrated Radio System}
\acro{MIT}{Massachusetts Institute of Technology}
\acro{MIT-LCS}{Massacchusetts Institute of Technology Laboratory for Computer Science}
\acro{ML}{Maximum Likelihood}
\acro{MLD}{Masking Level Difference}
\acro{MLE}{Maximum Likelihood Estimation}
\acro{ML-TDoA}{Maximum Likelihood Time Difference of Arrival}
\acro{MLMF}{Machine Learning with Multiple Features}
\acro{MLP}{Multi-layer Perceptron}
\acro{MLS}{Maximum Length Sequence\acroextra{ of pseudo random bits.}}
\acro{MMSE}{Minimum Mean Squared Error}
\acro{MMT}{Multiscale Median Transform}
\acro{MNRU}{Modulated Noise Reference Unit}
\acro{MOM}{Mean of Maximum}
\acro{MOS}{Mean Opinion Score}
\acro{MOS-LQO}{Mean Opinion Score - Listening Quality Objective}
\acro{MP}{Matching Pursuit}
\acro{MP3}{\acs{MPEG}-2 Audio Layer III}
\acro{MPEG}{Moving Picture Experts Group}
\acro{MRF}{Markov Random Field}
\acro{MRP}{Mouth Reference Point (cf. ITU-T Rec. P.64 1999)}
\acro{MRT}{Modified Rhyme Test}
\acro{MS}{Minimum Statistics}
\acro{MSB}{Most Significant Bit}
\acro{MSC}{Mean Square Coherence}
\acro{MSE}{Mean Square Error}
\acro{MSN}{Multiple Subscriber Number}
\acro{MSNR}{Maximum \acs{SNR}}
\acro{MTF}{Modulation Transfer Function}
\acro{MTM}{Modified Trimmed Mean}
\acro{MUSCLE}{MeasUred Single-CLustEr}
\acro{MUSHRA}{Multi-stimuli Test with Hidden Reference and Anchor}
\acro{MUSHRAR}{Multi-stimuli Test with Hidden Reference and Anchor for Reverberation}
\acro{MUSIC}{Multiple Signal Classification}
\acro{MVDR}{Minimum Variance Distortionless Response\acroextra{ beamformer}}
\acro{MWF}{Multi-channel Wiener Filter}
\acro{NAH}{Nearfield Acoustic Holography}
\acro{NB}{Narrowband}
\acro{NCM}{Normalized Coherence Metric}
\acro{NH}{Normal-Hearing}
\acro{NIRA}{Non-Intrusive Room Acoustics}
\acro{NISE}{Non-Intrusive \acs{SNR} estimation}
\acro{NISI}{Non-Intrusive Speech Intelligibility Estimation}
\acro{NISQ}{Non-Intrusive Speech Quality Estimation}
\acro{NIST}{National Institute of Standards and Technology}
\acro{NL}{Noise Level}
\acro{NLA}{Non-Linear Approximation}
\acro{NLMS}{Normalized Least Mean Squares\acroextra{ adaptive filter}}
\acro{NMCFLMS}{Normalized Multichannel Frequency Domain Least Mean Square}
\acro{NMF}{Non-negative Matrix Factorization}
\acro{NOISEX-92}{Database to Study the Effect of Additive Noise on Speech Recognition Systems}
\acro{NOIZEUS}{Noisy Speech Corpus for Evaluation of Speech Enhancement Algorithms}
\acro{NOSRMR}{Normalized Overall \acs{SRMR}}
\acro{NOS}{Number of Sources}
\acro{NOSRMR}{Normalized Overall \acs{SRMR}}
\acro{NPM}{Normalized Projection Misalignment}
\acro{NPV}{Negative Predictive Value}
\acro{NR}{Noise Reduction}
\acro{NS}{Noise Suppression}
\acro{NSRMR}{Normalised \acs{SRMR}}
\acro{NSRR}{Normalized Signal-to-Reverberation Ratio}
\acro{NSRMR}{Normalised \acs{SRMR}}
\acro{NSV}{Negative-Side Variance}
\acro{NTP}{Network Time Protocol}
\acro{NV}{Noise-Vocoding}
\acro{OBL}{Octave Band Level}
\acro{ODF}{Overdrive Factor}
\acro{Ofcom}{Office of Communications\acroextra{, the independent regulator and competition authority for the UK communications industries}}
\acro{OFDM}{Orthogonal Frequency Division Multiplexing}
\acro{OHC}{Outer Hair Cell}
\acro{OIM}{Objective Intelligibility Measure}
\acro{OLA}{Overlap-add}
\acro{OM-LSA}{Optimally Modified Log-Spectral Amplitude{ Estimator}}
\acro{OMP}{Orthogonal Matching Pursuit}
\acro{OSI}{Open Systems Interconnection}
\acro{OSPA}{Optimal Subpattern Assignment}
\acro{OSRMR}{Overall \acs{SRMR}}
\acro{PAB-SRMR}{Per acoustic band \acs{SRMR}}
\acro{PALM}{Passive Acoustic Localization and Mapping}
\acro{PAMS}{Perceptual Analysis Measurement System}
\acro{PARCOR}{Partial Correlation Coefficients}
\acro{PB}{Phonetically Balanced}
\acro{PBF}{Positive Boolean Function}
\acro{PCA}{Principal Components Analysis}
\acro{PCM}{Pulse-Code Modulation}
\acro{PDA}{Personal Digital Assistant}
\acro{PDA2}[PDA]{Probabilistic Data Association}
\acro{PDE}{Partial Differential Equation}
\acro{pdf}{Probability Density Function}
\acro{PDF}{Probability Density Function}
\acro{PE}{Parameter Estimation}
\acro{PEASS}{Perceptual Evaluation for Audio Source Separation}
\acro{PEFAC}{Pitch Estimation Filter with Amplitude Compression}
\acro{PESQ}{Perceptual Evaluation of Speech Quality}
\acro{PF}{Psychometric Function}
\acro{pgfl}[p.g.fl.]{Probability Generating Functional}
\acro{PHAT}{Phase Transform}
\acro{PHD}{Probability Hypothesis Density}
\acro{PIP}{Peak-Image Pairing}
\acro{PIV}{Pseudo-Intensity Vector}
\acro{PL}{Pseudo-likelihood}
\acro{PLC}{Packet Loss Concealment}
\acro{PLL}{Phase Locked Loop}
\acro{PLP}{Perceptual Linear Prediction}
\acro{PLR}{Perceived Level of Reverberation}
\acro{PM}{Phase Modulation}
\acro{PMF}{Probability Mass Function}
\acro{PMOS}{Predicted Mean Opinion Score}
\acro{PNN}{Parameterized Neural Network}
\acro{POLQA}{Perceptual Objective Listening Quality Analysis}
\acro{POTS}{Plain Old Telephone Service}
\acro{PPP}{Poisson Point Process}
\acro{PPS}{Pulse-Per-Second}
\acro{PPV}{Positive Predictive Value}
\acro{PReLU}{Parametric Rectified Linear Unit}
\acro{PRLM}{Phoneme Recognition and Language Modelling}
\acro{PRP}{Pair-wise Relative Phase-ratio}
\acro{PSD}{Power Spectral Density}
\acro{PSK}{Phase Shift Keying}
\acro{PSNR}{Peak Signal-to-Noise Ratio}
\acro{PSOLA}{Pitch Synchronous Overlap Add\acroextra{. A method of scaling a signal in time and pitch independently.}}
\acro{PSQM}{Perceptual Speech Quality Measurement}
\acro{PSSL}{Positional Sound Source Localization}
\acro{PSTN}{Public Switched Telephone Network}
\acro{PWD}{Plane-Wave Decomposition}
\acro{PTA}{Pure-Tone Audiology}
\acro{QAM}{Quadrature Amplitude Modulation}
\acro{QILAH}{Quadrisected Iterated Logarithm Amplitude Histogram\acroextra{ clipping detection method}}
\acro{QMF}{Quadrature Mirror Filter}
\acro{QoE}{Quality-of-Experience}
\acro{QoS}{Quality-of-Service}
\acro{QPSK}{Quadrature Phase Shift Keying}
\acro{RADAR}{RAdio Detection And Ranging}
\acro{RASTA}{Relative Spectral}
\acro{RASTA-PLP}{Relative Spectral Perceptual Linear Prediction}
\acro{RASTI}{Room Acoustics Speech Transmission Index\acroextra{ (superseded by STIPA)}}
\acro{RBM}{Restricted Boltzmann Machine}
\acro{RB-PHD}{Rao-Blackwellised \acs{PHD}}
\acro{RBPF}{Rao-Blackwellised Particle Filter}
\acro{RC}{Relative Criterion}
\acro{RDT}[$R_\textrm{DT}$]{Reverberation Decay Tail}
\acro{RDTF}{Relative Direct Transfer Function}
\acro{ReLU}{Rectified Linear Unit}
\acro{RF}{Radio Frequency}
\acro{RFI}{Radio Frequency Interference}
\acro{RFS}{Random Finite Set}
\acro{RHS}{Right-Hand Side}
\acro{RIP}{Restricted Isometry Property}
\acro{RIR}{Room Impulse Response}
\acro{RLS}{Recursive Least Squares\acroextra{ adaptive filter}}
\acro{RLSD}{Relative Log Spectral Distortion}
\acro{RMCLS}{Relaxed Multichannel Least Squares}
\acro{RMCLS-CIT}{Relaxed MultiChannel Least-Squares with Constrained Initial Taps}
\acro{RMS}{Root Mean Square}
\acro{RMSE}{Root Mean Square Error}
\acro{RNN}{Recurrent Neural Network}
\acro{ROC}{Receiver Operating Characteristic}
\acro{ROHC}{Robust Header Compression}
\acro{RP}{Received Pronunciation}
\acro{RPE}{Regular Pulse Excitation}
\acro{RRTF}{Relative Real-Time Factor}
\acro{RS}{Reverberation Suppression}
\acro{RSM}{Reflector Source Method}
\acro{RSS}{Received Signal Strength}
\acro{RSV}{Room Spectral Variance}
\acro{RT}{Reverberation Time}
\acro{RTAN}{Robustness to Additional Noise}
\acro{RTF}{Real-Time Factor}
\acro{RTF2}[RTF]{Room Transfer Function}
\acro{RTF3}[RTF]{Relative Transfer Function}
\acro{RV}{Random Variable}
\acro{RVP}{Recursive Vector Projection}
\acro{RWTH}{Rheinisch-Westf\"{a}lische Technische Hochschule}
\acro{S50}[$S_{50}$]{Intelligibility Function Gradient at the \ac{SRT}}
\acro{SA}{Spectral Amplitude}
\acro{SAA}{Synthetic Aperture Audio}
\acro{SAP}{Speech And Audio Processing}
\acro{SAR}{Speech-to-Artifact Ratio}
\acro{SAR2}[SAR]{Speaker Alternation Rate}
\acro{SAR3}[SAR]{Synthetic Aperture \ac{RADAR}}
\acro{SAS}{Synthetic Aperture \ac{SONAR}}
\acro{SB}{Subband}
\acro{SC-PHD}{Single Cluster \acs{PHD}}
\acro{SCAF}{Single Channel Adaptive Filter}
\acro{SCB}{Stochastic Codebook}
\acro{SCNR}{Single-Channel Noise Reduction}
\acro{SCOT}{Smoothed Coherence Transform}
\acro{SCNR}{Single-Channel Noise Reduction}
\acro{SC-PHD}{Single Cluster \acs{PHD}}
\acro{SCR}{Signal-to-Competition Ratio}
\acro{SCRIBE}{Spoken Corpus of British English}
\acro{SCT}{Speech Corruption Toolkit}
\acro{SCT2}{Short Conversation Test}
\acro{SD}{Semantic Differential}
\acro{SDB}{Superdirective Beamformer}
\acro{SDD}{Spectral Decay Distributions}
\acro{SDDMSB}{\acs{SDD} with Mel-spaced frequency bands}
\acro{SDDSA}{\acs{SDD} with Mel-spaced frequency bands and selective averaging}
\acro{SDDSA-G}{\acs{SDDSA} with Gerkmann noise estimator}
\acro{SDDSA-H}{\acs{SDDSA} with Hendriks noise estimator}
\acro{SDR}{Software Defined Radio}
\acro{SDR2}{Speech}
\acro{SDRAM}{Synchronous Dynamic Random Access Memory}
\acro{SDT}{Speech Description Taxonomy}
\acro{SEDF}{Subband \ac{EDF}}
\acro{SELD}{Sound Event Localization and Detection}
\acro{SEMG}{Surface Electromyography}
\acro{SFDR}{Spurious Free Dynamic Range}
\acro{SFM}{Single Feature with Mapping}
\acro{SFT}{Spherical Fourier Transform}
\acro{SH}{Spherical Harmonic}
\acro{SHD}{Spectral Harmonic Decomposition}
\acro{SHD2}[SHD]{Spherical Harmonic Domain}
\acro{SIE}{System Identification Error}
\acro{SII}{Speech Intelligibility Index}
\acro{SIImod}{Speech Intelligibility Index in the modulation domain}
\acro{SIM}{Subscriber Identity Module}
\acro{SIMO}{Single-Input-Multiple-Output}
\acro{SINAD}{Signal-to-Noise and Distortion Ratio}
\acro{SIP}{Session Initiation Protocol}
\acro{SIR}{Signal-to-Interference Ratio}
\acro{SIR2}[SIR]{Sequential Importance Resampling}
\acro{SIREAC}{Simulation of REal Acoustics\acroextra{ Software Tool}}
\acro{SIS}{Sequential Importance Sampling}
\acro{SL}{Speech Level}
\acro{SLAM}{Simultaneous Localization and Mapping}
\acro{SLLN}{Strong Law of Large Numbers}
\acro{SLM}{Sound Level Meter}
\acro{SMA}{Spherical Microphone Array}
\acro{SMARD}{Single- and Multichannel Audio Recordings Database}
\acro{SMERSH}{Spatiotemporal Averaging Method for Enhancement of Reverberant Speech}
\acro{SMIR}{Spherical Microphone array Impulse Response}
\acro{SMPTE}{Society of Motion Picture and Television Engineers}
\acro{SMS}{Short Message Service}
\acro{SNR}{Signal-to-Noise Ratio}
\acro{SNR2}[SNR]{Speech-to-Noise Ratio}
\acro{SNT}{Subspace Noise Tracking\acroextra{ algorithm}}
\acro{SOBM}{STOI-optimal Binary Mask}
\acro{SONAR}{SOund Navigation And Ranging}
\acro{SOLA}{Synchronous Overlap Add\acroextra{. A method of scaling a signal in time and pitch independently.}}
\acro{SPC}{Specificity}
\acro{SPEECON}{Speech Databases for Consumer Devices}
\acro{SPHERE}{NIST SPeech Header REsources\acroextra{ software with embedded Shorten Compression}}
\acro{SPIN}{Speech Perception In Noise}
\acro{SPL}{Sound Pressure Level}
\acro{SPP}{Speech Presence Probability}
\acro{SPQA}{Speech Quality Assurance Package}
\acro{SQNR}{Signal-to-Quantization Noise Ratio}
\acro{SR}{Sparse Representation}
\acro{SR2}[SR]{Spectral Rotation}
\acro{SRA}{Statistical Room Acoustics}
\acro{SRI}{SRI International\acroextra{. Formerly Standford Research Institute}}
\acro{SRMR}{Speech-to-Reverberation Modulation Energy Ratio}
\acro{SRP}{Steered Response Power}
\acro{SRP-PHAT}{Steered Response Power with Phase Transform}
\acro{SRP-TDE}{Steered Response Power with Time Delay Estimation}
\acro{SRR}{Signal-to-Reverberation Ratio}
\acro{SRT}{Speech Reception Threshold\acroextra{ (also known as Speech Recognition Threshold)}}
\acro{SS}{Spectral Subtraction}
\acro{SSB}{Single Side-Band}
\acro{SSI}{Synthetic Sentence Identification}
\acro{SSL}{Sound Source Localization}
\acro{SSN2}[SSN]{Simultaneous Switching Noise}
\acro{SSN}{Speech-Shaped Noise}
\acro{SSNR}{Segmental \acs{SNR}}
\acro{SSOBM}{Stochastic \acs{SOBM}}
\acro{SSRR}{Segmental Signal-to-Reverberation Ratio}
\acro{SSW}{Staggered Spondaic Word}
\acro{STFT}{Short Time Fourier Transform}
\acro{STI}{Speech Transmission Index}
\acro{STIPA}{Speech Transmission Index for Public Address Systems}
\acro{STITEL}{Speech Transmission Index for Telecommunication Systems}
\acro{STMI}{Spectro-Temporal Modulation Index}
\acro{STNR}{\acs{NIST}'s Speech-to-Noise Ratio\acroextra{ Estimation Algorithm}}
\acro{STOI}{Short-Time Objective Intelligibility\acroextra{ measure}}
\acro{STQ}{Speech Processing, Transmission and Quality Aspects}
\acro{STSA}{Short Time Spectral Analysis}
\acro{STSA1}[STSA]{Short Time Spectral Amplitude}
\acro{SUS}{Semantically Unpredictable Sentences}
\acro{SVD}{Singular Value Decomposition}
\acro{SVM}{Support Vector Machine}
\acro{SWSOBM}{Stochastic \acs{WSOBM}}
\acro{T20}[$T_\textrm{20}$]{Reverberation Time\acroextra{ to decay by $20$ dB}}
\acro{T30}[$T_\textrm{30}$]{Reverberation Time\acroextra{ to decay by $30$ dB}}
\acro{T60}[$T_\textrm{60}$]{Reverberation Time\acroextra{ to decay by $60$ dB}}
\acro{TBM}{Target Binary Mask}
\acro{TDE}{Time Delay Estimation}
\acro{TDHS}{Time Domain Harmonic Scaling\acroextra{. A method of scaling a signal in time and pitch independently.}}
\acrodefplural{TDOA}[TDOAs]{Time-Differences-of-Arrival}
\acro{TDOA}{Time-Difference-of-Arrival}
\acrodefplural{TDOA}[TDOAs]{Time-Differences-of-Arrival}
\acro{TDT}{Tone Decay Test}
\acro{TF}{Time-Frequency}
\acro{TFS}{Temporal Fine Structure}
\acro{TFGM}{Time-Frequency Gain Modification\acroextra{. An approach to signal enhancement in which a signal is multiplied by a gain function in the time-frequency domain.}}
\acro{THD}{Total Harmonic Distortion}
\acro{TI}{Texas Instruments, Inc.}
\acro{TIMIT}{\acs{TI}-\acs{MIT} speech corpus}
\acro{TIPHON}{Telecommunication and Internet Protocol Harmonization Over Networks}
\acro{TLS}{Total Least-Squares}
\acro{TN}{True Negative}
\acro{TNR}{True Negative Rate}
\acro{TOA}{Time-of-Arrival}
\acrodefplural{TOA}[TOAs]{Times-of-Arrival}
\acro{TOF}{Time-of-Flight}
\acro{TOSQA}{Telekom Objective Speech Quality Assessmentt}
\acro{TP}{True Positive}
\acro{TP2}[TP]{Trivial Pursuit}
\acro{TPCC}{Trivial Pursuit with Clipping Constraints}
\acro{TPR}{True Positive Rate}
\acro{TSE}{Taylor Series Expansion}
\acro{TVAR}{Time-varying Autoregression}
\acro{UAV}{Unmanned Aerial Vehicle}
\acro{UDP}{User Datagram Protocol}
\acro{UFRJ}{Federal University of Rio de Janeiro}
\acro{UHF}{Ultra High Frequency}
\acro{UKF}{Unscented Kalman Filter}
\acro{ULA}{Uniform Linear Array}
\acro{ULF}{Ultra Low Frequency}
\acro{UMTS}{Universal Mobile Telecommunications Service}
\acro{US}{United States}
\acro{UTBM}{Universal Target Binary Mask}
\acro{VAD}{Voice Activity Detector}
\acro{VBR}{Variable Bit-Rate}
\acro{VCV}{Vowel-Consonant-Vowel}
\acro{VRD}{Variance of Decay-rates}
\acro{VGC}{Voice Grade Channel}
\acro{vMF}{von Mises-Fisher}
\acro{VoIP}{Voice Over Internet Protocol}
\acro{VRT}{Vlaamse Radio- en Televisieomroeporganisatie\acroextra{. (Flemish Radio and Television Broadcasting Organization)}}
\acro{VSELP}{Vector Sum-excited Linear Prediction}
\acro{VST}{Virtual Studio Technology\acroextra{. An interface standard developed by Steinberg for adding plugins to an audio editor.}}
\acro{WADA}{Waveform Amplitude Distribution Analysis}
\acro{WASN}{Wireless Acoustic Sensor Network}
\acrodefplural{WASN}[WASNs]{Wireless Acoustic Sensor Networks}
\acro{WASPAA}{Workshop on Applications of Signal Processing to Audio and Acoustics}
\acro{WAV}{Waveform Audio File Format}
\acro{WAVE}{Waveform Audio File Format}
\acro{WB}{Wideband}
\acro{WER}{Word Error Rate}
\acro{WGN}{White Gaussian Noise}
\acro{WLAN}{Wireless \acs{LAN}}
\acro{WLLN}{Weak Law of Large Numbers}
\acro{WM}{Working Memory}
\acro{WMA}{Windows Media Audio}
\acro{WNG}{White Noise Gain}
\acro{WSS}{Weighted Spectral Slope}
\acro{WSOBM}{Weighted \acs{SOBM}}
\acro{WSTOI}{Weighted \acs{STOI}}
\acro{ZOS}{Zero-Order Statistics}

%% file: introduction.tex
\section{Introduction}

Most signals, such as audio and images, contain metadata. Metadata can be signal-based, which describes quantitative properties of the signal, such as its sampling rate, as well as semantic, which describes, for example, contextual properties. In speech processing, semantic metadata could consist of the speaker's language or gender. Whether signal-based or semantic, including metadata as a secondary input into neural network models may provide relevant information which would translate into an economy of training time, model parameters and flexibility. However, metadata typically has a different dimensionality than the input signals, making its incorporation into those models not trivial. 

The main focus of this paper is to study the effectiveness of schemes to process signals and exploit metadata jointly using neural network models. We focus on the task of \acf{SSL} \cite{So2011} using distributed microphone arrays to demonstrate the effectiveness of our proposed approach. In the context of \ac{SSL}, relevant metadata which is exploited by classical methods is the microphone positions, which can be acquired by manual measurement or using self-calibration \cite{Haddad2017} methods. Other relevant metadata is the room dimensions and its reverberation time.  

SSL refers to the task of estimating the spatial location of a sound source, such as a human talker or a loudspeaker. In this scenario, metadata refers to properties of the acoustic scene such as the coordinates of microphones, dimensions of the room and, the reflection coefficient of the walls. SSL has many applications, including noise reduction and speech enhancement \cite{brandstein2001microphone}, camera steering \cite{Wang1997} and acoustic Simultaneous Localization and Mapping (SLAM) \cite{Evers2018c}. In turn, distributed microphone arrays have
become an active research topic in the signal processing community due to their versatility. Such arrays may be composed of multiple network-connected devices, including everyday devices such as
cell phones, smart assistants, and laptops, for example. The array and the constituent devices may be
configured as a Wireless Acoustic Sensor Network (WASN) \cite{Bertrand2011}.

% Problem statement
\ac{SSL} approaches may be divided into classical signal processing-based and data-driven neural network-based methods. By explicitly exploiting metadata describing microphone positions and room dimensions, classical approaches may be applied to different rooms and microphone configurations. Conversely, neural network approaches have recently achieved state of the art results for source localization \cite{Adavanne2018, He2018, Vera-Diaz2018}, at the expense of requiring one network to be trained for every microphone topology. One reason current neural approaches do not incorporate the microphones' positional information is that the microphones' signal and positional data are very different from one another in nature and dimension. 

Previous work which discusses the joint processing of signals and metadata is \cite{Baldi2016}, where a single input neural network is used to process metadata in conjunction with a low-dimensional physical signal. However, unlike our work, the method of \cite{Baldi2016} is restricted to multilayer perceptron architectures and one-dimensional input and metadata, limiting its application in practical scenarios.

Another related field is multimodal fusion \cite{Atrey2010, Baltrusaitis2019}, although this is usually concerned with learning representations using two types of signals, such as audio-visual data. Simultaneously processing signals and metadata have also been explored using non-neural models for sound source separation \cite{Ozerov2012}, where metadata consists of information about the type of sound (speech, music) and how the sources were mixed. However, none of the existing work discusses effective schemes for incorporating and evaluating signals and metadata of different dimensionality.\\

\begin{figure*}
    \begin{center}
        \includegraphics{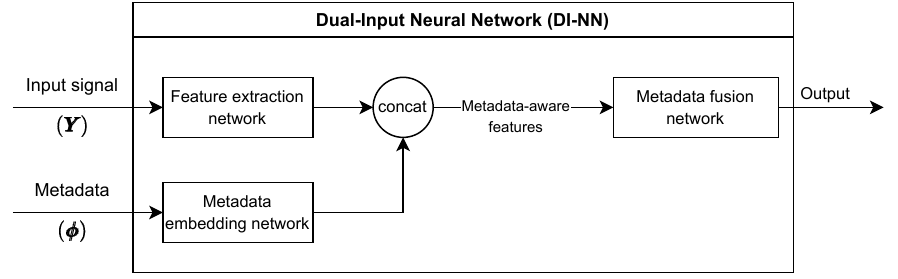}
    \end{center}
    \caption{Overview of the \ac{DI-NN} approach.}
    \label{fig:di-nn}
\end{figure*}

Our main contribution is the DI-NN neural network architecture, which is capable of processing high-dimensional signals, namely spectrograms, along with a relevant metadata vector of lower dimensionality. An overview diagram of our approach is shown in Fig.~\ref{fig:di-nn}, which will be discussed in \autoref{sec:dinn}. We compare our method against three baselines for the task of \ac{PSSL}, namely, a metadata-unaware \acf{CRNN}, a metadata-aware classical signal processing approach, as well as an alternate metadata-aware neural network. Our proposed method is able to outperform all baselines by a large margin in realistic scenarios. In contrast to previous approaches \cite{Sundar2020, Vera-Diaz2018}, our network dispenses with the need for training a network for each scenario, broadening our method's applicability.

This work continues as follows. In \autoref{sec:ssl}, an overview of neural and non-neural \ac{SSL} methods will be discussed. The approach for training our proposed \ac{DI-NN} for \ac{SSL} is described together with several baseline methods in \autoref{sec:architecture}. In \autoref{sec:experimentation}, the experiments comparing our approach with the baselines using multiple datasets are described. Finally, results and conclusions are drawn in \autoref{sec:conclusion}.

%% file: prior_art.tex
\section{Prior art on sound source localization}
\label{sec:ssl}

\subsection{Neural-based methods}

In recent years, deep neural networks have been widely adopted for the task of sound source localization. The various approaches differ in the input features used, the network architectures and output strategies. Most studies focus on the task of \ac{DOA} estimation, i.e., estimating the angle between the propagation direction of the acoustic wavefront due to the source and a reference axis of the array.

Practicioners have experimented with many types of neural input features, such as the raw audio samples of the microphone signals \cite{Vera-Diaz2018}, their frequency-domain representation through the \acf{STFT} \cite{Anonymous2022}, their cross-spectra \cite{Xue2020} or cross-correlation \cite{He2018}. Multiple architectures have been also tested, including the \ac{MLP} \cite{He2018}, \acp{CNN} \cite{Chakrabarty2017} and residual networks \cite{Yalta2017}. In this work, we focus on the \acf{CRNN} architecture, which has received widespread adoption in the field \cite{Krause2021, Adavanne2018, Perotin2019}. Finally, approaches differ in terms of the network's output strategy. While regression-based approaches directly estimate the source's coordinates, classification based-approaches discretize the source locations to a grid of available positions. We refer to \cite{Perotin2019a} for a discussion on the merits of both approaches. We also refer the reader to a substantial survey of neural SSL papers \cite{Grumiaux2021}.

In this paper, we focus on the task of estimating the absolute Cartesian coordinates of the source, which we shall refer to as \acf{PSSL}, and has applications in robot navigation \cite{Evers2018c} and noise reduction \cite{Taseska2013a}. The \ac{PSSL} task has been much less studied using neural methods. To the best of our knowledge, only \cite{Sundar2020} and \cite{Vera-Diaz2018} focus on \ac{PSSL}. However, both these approaches only work for the same room with fixed relative microphone positions. We believe this shortage of studies to be at least in part due to the lack of an architecture capable of incorporating the scene's metadata, which is addressed by our proposed \ac{DI-NN}. We also refer to the recent L3DAS22 challenge \cite{Guizzo2022}, where practitioners were invited to develop 3D \ac{PSSL} algorithms for a realistic office environment containing a pair of microphone arrays.

\subsection{Classical signal processing methods}

Classical approaches to \ac{SSL} have been widely studied within the signal processing community. In \ac{PSSL} approaches, the source's coordinates are estimated using a model involving signal processing, physics and geometry. By measuring differences in the microphone signals' amplitudes and phases, distance metrics between the microphones and source can be estimated. These estimates can in turn be combined to estimate the source's coordinates \cite{So2011}. Besides the microphones' signals, the positions of microphones are usually needed for the position of the source to be estimated. Available approaches for \ac{SSL} may be classified as delay-based \cite{Gustafsson2003a, So2011}, energy-based \cite{Li2003b, Liu2007a}, subspace-based  \cite{Schmidt1986a} and beamforming-based \cite{Dmochowski2007e, Lebarbenchon2018} approaches. We shall focus on delay-based approaches and will provide background for our baseline method.

Delay-based \ac{SSL} methods usually rely on computing the \ac{TDOA} between each microphone pair within the system, which corresponds to the difference in time taken for the source signal to propagate to different microphones. The locus of candidate source positions with the same \ac{TDOA} with respect to a microphone pair is, when considering planar coordinates, a hyperbola \cite{So2011, Gustafsson2003a}. The source is located at the intersection of the hyperbolae defined by all microphone pairs.
The multiple \acp{TDOA} can be combined using a \acf{LS} framework \cite{Huang2001c}, or using a \ac{ML} approach if some noise properties of the system are known \cite{So2011}. In general, \acp{TDOA} are estimated using cross-correlation based methods such as \ac{GCC-PHAT} \cite{Knapp1976b}, which are shown to be somewhat robust to reflections produced in the room due to, for example, the walls, ceiling and furniture, i.e. reverberation \cite{Zhang2008a}.

%% file: method.tex
\section{Method} \label{sec:architecture}

\subsection{Signal model and scope of this work} \label{sec:scope}
Our scope is restricted to the localization of a static source at the planar coordinates $\pmb{p}_s = [p_s^x, p_s^y]^T$. The source emits an intermittent signal $s(t)$ at time $t$. In our experiments, $s(t)$ may consist of \ac{WGN} as well as of speech utterances.
Also, $M$ static microphones with known positions are present in the room, each placed at coordinates $\pmb{m}_i = [m_i^x, m_i^y]^T$. Both source and microphones are enclosed in a room of planar dimensions $\pmb{d}=[d^{x}, d^{y}]^T$. The amount of reverberation in the room is modeled by its reverberation time $r$, a measure of the amount of time it takes for a sound to decay by 60\,dB from its original level. The signal $y_i$ received at microphone $i$ is
\begin{equation} \label{eq:received_signal}
    y_i(t) = a_i s(t - \tau_i) + \epsilon_i(t) \;.
\end{equation}
In \eqref{eq:received_signal}, $a_i$ is a scaling factor representing the attenuation suffered by the wave propagating from $\pmb{p}_s$ to $\pmb{m}_i$. We assume that the gains between the microphones are approximately calibrated, although we show in \autoref{sec:real_recordings} that our method is robust to uncalibrated microphones of the same kind. $\tau_i$ is the time taken for a sound wave to propagate from the source to microphone $i$, and $\epsilon_i(t)$ models the noise. We assume $\tau_i$ to be equal to $\Vert \pmb{m}_i - \pmb{p}_s \Vert_2 /c$, where $\Vert \pmb{m}_i - \pmb{p}_s\Vert_2$ is the Euclidean distance between the source and the microphone located at $\pmb{m}_i$, $c$ is the speed of sound and $\Vert \cdot \Vert_2$ represents the $L_2$-norm. 

We also define $\pmb{y}(t) = [y_1(t), \dotsc, y_M(t)]^T$ as the vector containing all microphone signals at discrete time index $t$. The \acf{STFT} of $y_i(t)$ is $Y_i(\ell,f)$, for frequency $f$ and time frame $\ell$, and $\pmb{Y}(\ell,f) = [Y_1(\ell,f), \dotsc, Y_M(\ell,f)]^T$. The \ac{STFT} \cite{Allen1977c} represents the frequency content of a signal over time, and is a widely used feature for source localization using neural networks \cite{Krause2021, Anonymous2022}. Figure \ref{fig:detailed-architecture} shows the magnitude representation of $\pmb{Y}$ at the input.

Finally, the \textit{metadata vector} $\pmb{\phi} \in \mathbb{R}^{N_{\phi}}$ is the concatenation of the coordinates of the microphones, the room dimensions and reverberation time, as shown in Fig. \ref{fig:detailed-architecture}. We chose the three aforementioned types of metadata as the room dimensions and microphone coordinates are explicitly exploited in classical localization methods such as the \ac{LS}. Furthermore, we included the reverberation time as an additional metadata to verify whether its knowledge can reduce the detrimental effect of reverberation in localization methods. However, other metadata could have been exploited such as the energy ratio between the microphone signals, or the absoption coefficients of the walls.

\input{dinn}
\input{efnn}

\input{least_squares_localization}

%% file: dinn.tex
\subsection{Proposed method: Dual input neural network} \label{sec:dinn}
% Show a detailed architecture of the network.

% Feature extraction network
Our proposed \ac{DI-NN} architecture is comprised of two neural networks, a \textbf{feature extraction network} and a \textbf{metadata fusion network} as can be seen in Fig.~\ref{fig:di-nn}. An additional third network, called the \textbf{metadata embedding network} is also used in the alternative DI-NN-Embedding network, which will be presented in \autoref{sec:di-nn-embedding} .

% input
The input of the network consists of the \ac{STFT}of the microphone signals as defined in \autoref{sec:scope}. Instead of using the complex representation generated by the \ac{STFT}, we split the real and imaginary parts of the \ac{STFT} $\pmb{Y}$ use them as separate channels as in \cite{Krause2021}, giving rise to $2*M$ input channels. The role of the feature extraction network is to transform this high dimensional tensor into a one dimensional feature vector which compactly represents relevant information for the task in hand. In our experiments, we adopt a \ac{CRNN} \cite{Choi2017} as our feature extraction network, due to its wide adoption for \ac{SSL} \cite{Adavanne2018, Perotin2019, Cao2019}.

This metadata-unaware vector is then concatenated to the available metadata, thus creating a metadata-aware feature vector. For our application, the metadata is a one-dimensional vector consisting of the positions of the microphones, the dimensions of the room, and its reverberation time. This metadata-aware feature vector is then fed to a metadata fusion network, whose role is to merge the metadata and feature vector to produce the result. In our experiments, we adopt a two-layer \acf{FC-NN} which maps the metadata-aware features to a two dimensional vector corresponding to the estimated coordinates of the source.

\begin{figure*}
    \begin{center}
        \includegraphics[width=0.9\textwidth]{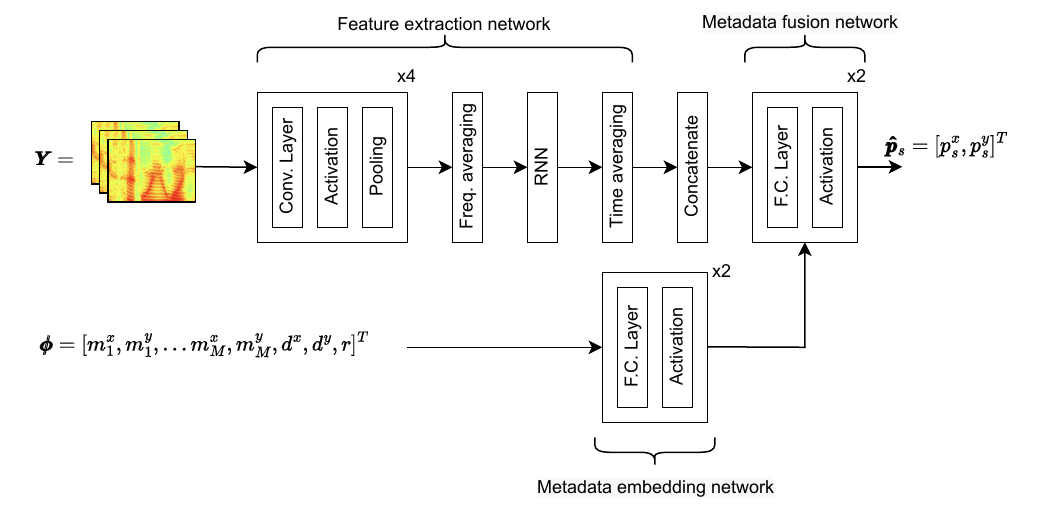}
    \end{center}
    \caption{Detailed DI-NN architecture for the task of \ac{PSSL}.}
    \label{fig:detailed-architecture}
\end{figure*}

Our feature extractor \ac{CRNN} is divided into two sequential sub-networks: a \ac{CNN} block, responsible for extracting local patterns from the input data and a \ac{RNN}, responsible for combining these pattens into global, time-independent features. A diagram representing the components of the DI-NN network is shown in Fig. \ref{fig:detailed-architecture}.

The convolutional block receives a tensor of shape $(M, L, F)$ representing a multi-channel complex \ac{STFT}, where $M$ represents the number of audio channels, $L$ represents the number of time frames generated by the \ac{STFT}, and $F$ is the number of frequency bins used. The role of this block is two-fold: firstly, to combine local information across all microphone channels, and secondly to reduce the dimensionality of the data to make it more tractable for the \ac{RNN} layer.

The convolutional block consists of four sequential layers, where each performs three sequential operations. Firstly, a set of $K$ convolutional filters is applied to the input signal, resulting in $K$ output channels. Secondly, a non-linear activation function is applied to the result. Finally, an average pooling operation is applied to the width and height of the activations, generating an output of reduced size. After passing the input through the four convolutional layers, we perform a global average pooling operation across all frequencies, generating a two-dimensional output matrix. 

After the convolutional block, the resulting matrix serves as input to a bidirectional, gated recurrent unit neural network (GRU-RNN) \cite{Chung2014a}. As sound may not be present throughout the whole duration of the audio signal, such as during speech pauses, the \ac{RNN} is important for propagating location information to silent time-steps. After this network, we reduce the dimensions of the features once again by performing average pooling on the time dimension, resulting in a vector of time-independent features. 

%\subsection{Output layer and loss function}
The output of the feature extraction network are then concatenated to the available metadata and serve as input to the metadata fusion network. This network consists of a set of two fully connected layers which map the metadata-aware features to a two-dimensional vector corresponding to the estimated cartesian coordinates of the active source. We jointly train both networks using the same loss function, defined as the $L_1$-norm or the sum of the absolute error between the network's estimate of the source coordinates $\pmb{\hat{p}}_s$ and the target $\pmb{p}_s$, given by 
\begin{equation} \label{eq:loss}
    \mathcal{L}(\pmb{p}_s, \pmb{\hat{p}}_s) = |\pmb{p}_s - \pmb{\hat{p}}_s| \;.
\end{equation}
We also considered using the more common squared error loss. Although both losses yielded similar results in our experiments, we chose the absolute error for its easier interpretability, since it corresponds to the distance in metres between target and estimated coordinates. 
% Loss
%We define the \ac{DOA} between the microphone array and the source as the counter-clockwise angle between the microphone and source vectors. 

%% file: efnn.tex
\subsection{DI-NN-Embedding} \label{sec:di-nn-embedding}

To test whether it is advantageous to process the metadata before combining it with the microphone features, we also propose a variant of the \ac{DI-NN} model, where the metadata $\pmb{\phi}$ is processed by a \textit{metadata embedding network} to produce an embedding, which is then concatenated to the microphone features. This network is represented by the \textit{metadata embedding network} block in \autoref{fig:di-nn}.

%% file: least_squares_localization.tex
\subsection{Baseline: Least-squares based source localization} \label{sec:ls}
%\vspace{-0.2cm}
\begin{figure*}
    \begin{center}
        \includegraphics[width=0.9\textwidth]{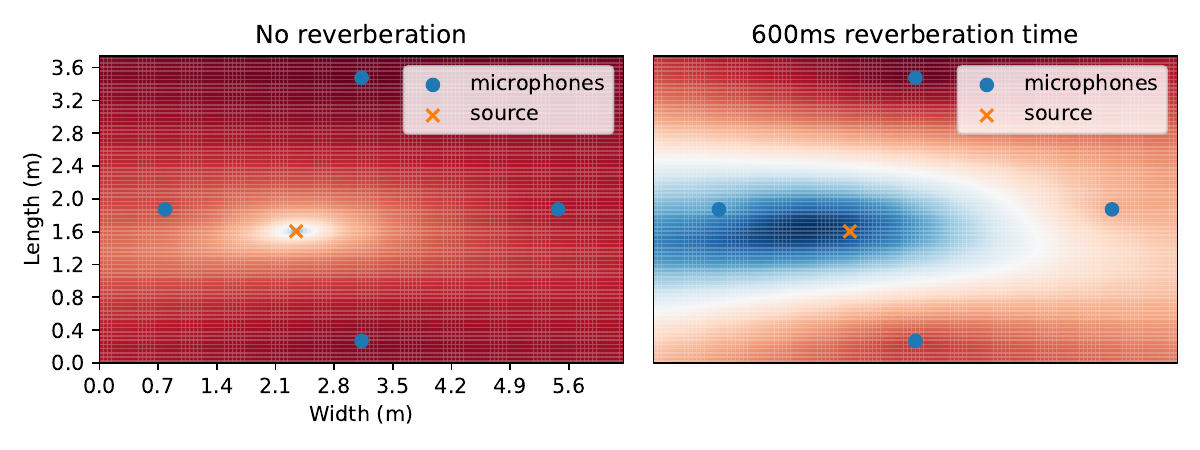}
    \end{center}
    \caption{Error grid produced by the LS algorithm for an anechoic and a reverberant room of the same dimensions and microphone coordinates.}
    \label{fig:ls_grid}
\end{figure*}

Our final comparative baseline is the \acf{LS} algorithm \cite{So2011} which uses the signal model defined in \autoref{sec:scope}. We provide an overview of the algorithm below. We define the \textit{theoretical} \ac{TDOA} between microphones $i$ and $j$ with respect to the source coordinates $\pmb{p}_s$ as
\begin{equation} \label{eq:tdoa}
    \tau_{ij}(\pmb{p}_s) \triangleq \frac{\Vert \pmb{m}_i - \pmb{p}_s \Vert_2 - \Vert \pmb{m}_j - \pmb{p}_s \Vert_2}{c} \;,
\end{equation}
where $c$ is the speed of sound. Next, the \textit{measured} \ac{TDOA} between microphones $\hat{\tau}_{ij}$ is estimated from the cross-correlation peak between the received signals according to
\begin{equation} \label{eq:gcc}
    \hat{\tau}_{ij} \triangleq \argmax_t (\mathcal{C}(t; y_i, y_j)) \;,
\end{equation}
where $\mathcal{C}$ denotes the cross-correlation operator, usually computed in the frequency domain using the \ac{GCC-PHAT} algorithm \cite{Knapp1976b}. We then aggregate the total error for all microphone pairs using
\begin{equation} \label{eq:error_ij}
    E(\pmb{p}_s) \triangleq \sum_{i=1}^{m} \sum_{j \neq i} E_{ij}(\pmb{p}_s) \;,
\end{equation}
where $ E_{ij}(\pmb{p}_s) \triangleq |\tau_{ij}(\pmb{p}_s) - \hat{\tau}_{ij}|^2 $ is the squared difference between the theoretical and measured \ac{TDOA} of each microphone pair in \eqref{eq:tdoa} and \eqref{eq:gcc}, respectively. To estimate the source's location, we compute the values of $E$ for a set of candidate locations $\pmb{p}_s$ within the room. In the absence of noise and reverberation, the location with the minimum error corresponds to the true position of the source \cite{So2011}. Figure \ref{fig:ls_grid} shows the heatmaps or error grids generated using the LS algorithm in an anechoic and a reverberant room. The position of the source is estimated by selecting the positions that minimize the total error,
\begin{equation} \label{eq:source_estimation}
    \hat{\pmb{p}}_s = \argmin_{p_s} E(\pmb{p}_s) \;.
\end{equation}
Figure \ref{fig:ls_grid} illustrates the limitations of the LS algorithm when the reverberation time is large. The two figures show the results of our algorithm for two simulations, where one source and four microphones are placed in a room with the same dimensions. When the room is simulated to be anechoic, i.e., all the reflections are absorbed, the algorithm produces a sharp blue peak in the heatmap. Conversely, when the simulated room is reverberant, the peak becomes much more dispersed. An explanation for this is that the model used by the \ac{LS} method assumes anechoic propagation between the source and microphones, i.e., no reflections are assumed. Conversely, we will show that the DI-NN model is able to localize sources in reverberant environments, as it is trained using a reverberant dataset. A study conducted in \cite{Bistafa2000} shows that speech inteligibility is maximized in rooms with a reverberation time between 0.4 and 0.5~ms, therefore limiting the practical application of the \ac{LS} method on those environments.

%% file: experimentation.tex
\section{Experimentation} \label{sec:experimentation}

This section describes our experiments with \acp{DI-NN} with three \ac{SSL} datasets representing scenarios of varying difficulties. For each dataset, our approach is compared to two other methods. The first method is a \ac{CRNN} with the same architecture but without using the available metadata, i.e., without the ``Concatenate" block in Fig. \ref{fig:detailed-architecture}. By comparing this network's performance to the \ac{DI-NN}, we can see the performance gains of our proposed method. The second comparative method is the classical \ac{LS} source localization method described in \autoref{sec:ls}. The experiments will be described below.

All of our experiments consisted of randomly placing one source and four microphones within a room. The height of the microphones, source and room were fixed for all experiments. For each experiment, the goal of the proposed method and baselines was to estimate the planar coordinates of the source within the room using a one-second multichannel audio signal as well as the positions of the microphones. We emphasize that the training and testing samples do not overlap, and hence demonstrate our method's effectiveness for handling unseen scenes and metadata. We refer the reader to Appendix A for a discussion on the independence of our datasets.

% Code
To simulate sound propagation in a reverberant room, we used the image source method \cite{Allen1979a} implemented by the Pyroomacoustics Python library (MIT license) \cite{Scheibler2018a}. We trained our neural networks using PyTorch (BSD license) \cite{Paszke2019a} along with the PyTorch Lightning (Apache 2.0 license) library \cite{Falcon2019a}. The models were trained using a single NVIDIA P100 GPU with 16~GB of RAM memory. The configuration of our experiments is managed using the Hydra\, (MIT license) library \cite{Yadan2019}. We release the code used for generating the data and training the networks on GitHub \footnote{Code: \url{https://github.com/egrinstein/di_nn}}, as well as a Kaggle notebook\,\footnote{Demo notebook: \\ \url{https://kaggle.com/code/egrinstein/di-nn-training-notebook}} to allow reproduction of the experiments without the need for any local software installation. The hyperparameters used for training the proposed method and baselines are shown in Table~\ref{table:hyperparams}.

\begin{table*}[ht]
    \caption{Hyperparameters} % title of Table
    \centering % used for centering table
    \begin{tabular}{c c} % centered columns
    \hline\hline %inserts double horizontal lines
    Parameter & Value \\ [0.5ex] % inserts table
    %heading
    \hline % inserts single horizontal line
    Num. parameters (DI-NN) & 3.5M \\
    Num. conv. kernels & 64, 128, 256, 512 \\    
    Conv. kernel size & 2x2 \\
    Conv. layer pooling size & 2x2 \\
    GRU output size & 256 \\
    Metadata fusion net. layer out. sizes & $512 + N_{\phi}$, $2$ \\
    Metadata embedding layer out. sizes & $2 N_{\phi}$, $N_{\phi}$  \\
    Activation func. last layer & None \\
    Activation func. other layers & \ac{ReLU} \\
    Num. \ac{DFT} bins (for \ac{STFT}) & 1024 \\
    \ac{DFT} hop length (for \ac{STFT}) & 512 \\
    Input duration & 0.5 secs. \\
    Sampling rate & 16kHz \\
    Grid resolution of LS method & 2~cm  \\ [0.5ex] % [1ex] adds vertical space
    Learning rate & 0.0005  \\
    Batch size & 32 \\
    Num. epochs & 40 \\ % inserting body of the table
    Batch normalization \cite{Ioffe2015} & Only after conv. layers \\
    Optimizer & Adam \cite{Kingma2017}   \\
    
    \hline %inserts single line
    \end{tabular}
    \label{table:hyperparams} % is used to refer this table in the text
\end{table*}

\begin{figure*}
    \begin{center}
        \includegraphics{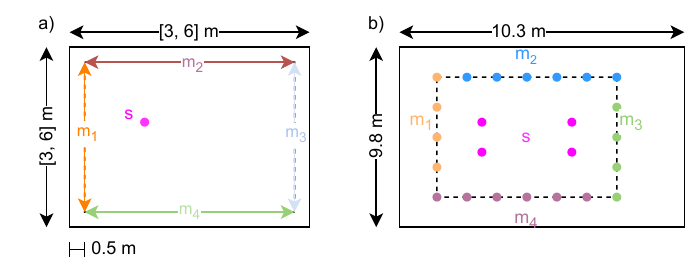}
    \end{center}
  \caption{Experimental setup. (a) For the anechoic and reverberant simulations, each of the four microphones $m_i$ is placed on a random point along the the coloured arrows, while the source $s$ is randomly placed on a point within the rectangle defined by them. (b) The sampling procedure for \autoref{sec:real_recordings}, where positions of the microphones and source are randomly drawn from each differently coloured set of points.}
  \label{fig:dataset-generation}
\end{figure*}

\subsection{Simulated anechoic rooms} \label{sec:simulated_anechoic}
The goal of this experiment is to evaluate the performance of the \ac{DI-NN} and baselines in multiple rooms and microphone positions in the absence of reverberation. Our dataset generation procedure is shown in Fig. \ref{fig:dataset-generation}a. For each dataset sample, we randomly select two numbers from a uniform distribution in the interval [3, 6]\,m representing the room's width and length. The height of the rooms is fixed at 3\,m. Next, we randomly place one microphone along a line segment 0.5\,m away and parallel to each room's walls. We chose to place the microphones close to the wall as a simplified localization scenario, as our main goal is to test the effectiveness of our metadata fusion procedure. Nonetheless, this scenario is realistic in the context of smart rooms, where the microphones are usually placed in or near the room's walls.

Finally, the source is randomly placed in the room, following a uniform distribution while respecting a minimum margin of 0.5\,m from the walls. In this experiment, the source signal is \ac{WGN}, and 30\,dB \ac{SNR} sensor noise, simulated using \ac{WGN}, is also added to each microphone. A dataset of 15,000 samples is generated, from which 10,000 samples are used for training, 2,500 for validation, and 2,500 for testing.

\subsection{Simulated reverberant rooms} \label{sec:simulated_reverb}
The data for the simulated reverberant rooms experiment is generated similarly to the anechoic experiment. However, instead of simulating sound propagation in an anechoic environment, each dataset sample is randomly assigned a reverberation time value for its corresponding room from a uniform distribution within the range of [0.3 -- 0.6]\,s. This value is used to simulate reverberation using the image source method \cite{Allen1979a}. For the source signal, we use speech recordings from the VCTK corpus \cite{Yamagishi2019a}. The number of training, testing and validation samples is same as in the above section.

\subsection{Real recordings} \label{sec:real_recordings}
For this experiment, instead of simulations, we use measurements from the LibriAdhoc40 dataset \cite{Guan2021} (GPL3 license). The signals were recorded in a highly reverberant room containing a grid of forty microphones and a single loudspeaker, which was placed in one of four available locations. The microphones recorded speech sentences taken from the Librispeech \cite{Panayotov2015a} corpus, which were played back through the loudspeaker. The reverberation time measured by the dataset authors was of approximately 900~ms.

To generate each dataset sample, we subselect four of the forty available microphones. We restrict our microphone selection to the outermost microphones of the grid, where one microphone per side is selected. A visual explanation of our microphone selection procedure is provided in Fig.~\ref{fig:dataset-generation}b. There are four available positions for the microphones near each of the west and east walls and seven positions near each of the north and south walls. Furthermore, there are four available source positions. There are, therefore, $4\times 4 \times 7\times 7 \times 4 = $~3,136 source/microphone combinations available for selection. Finally, we randomly select four speech utterances for each combination, resulting in a dataset of 12,544 samples. We use 50\% of those combinations for training, 25\% for validation and 25\% for testing. To create the training dataset for this experiment, we augment the aforementioned training split with the training data of the reverberant dataset described in \autoref{sec:simulated_reverb}, resulting in a dataset consisting of 10,000 $+$ 6,272 $=$ 16,272 signals. %The goal of this augmentation is to require the learned model to be able not only to perform source localization in the real room but also  

\subsection{Metadata sensitivity study} \label{sec:sensitivity}
In practical scenarios, the metadata, e.g., microphone coordinates and room reverberation time in \ac{PSSL}, are uncertain because they are typically estimated or measured. To investigate the robustness of our approach to such uncertainties, we conducted a sensitivity study using the test dataset in \autoref{sec:simulated_reverb}. We modify the dataset by introducing different levels of perturbations to the input metadata, followed by a computation of the mean localization error for each level using the model trained on \autoref{sec:simulated_reverb}.

Our first three studies consist of perturbing the microphone coordinates of the testing dataset with increasing levels of random Gaussian noise. The reported precision of microphone coordinates measured optically is under a millimeter \cite{Aurand2017a}. Conversely, when these are estimated using self-localization algorithms, the reported errors are under 7\,cm \cite{Gaubitch2013, Pertila2012}. We therefore choose the standard deviation levels of the introduced noise to 1, 10 and 50\,cm. In our fourth study, we introduced random Gaussian noise to the reverberation time with a standard deviation of 200\,ms, based on reported errors obtained on reverberation estimation procedures \cite{Gamper2018a, Lopez2021}.

\subsection{Metadata relevance study} \label{sec:relevance}

To quantify the contribution of each metadata category to the improvement in localization performance, we conducted a metadata relevance study where we trained the \ac{DI-NN} network using six different combinations of the microphone positions, room dimensions and reverberation time. The results are summarized in \autoref{table:relevance}.

%% file: conclusion.tex
\section{Results and discussions} \label{sec:conclusion}

% maybe you could use paragraph instead of subsection
\subsection{Results} \label{sec:results}

\begin{figure}
    \begin{center}
        \includegraphics[width=0.9\columnwidth]{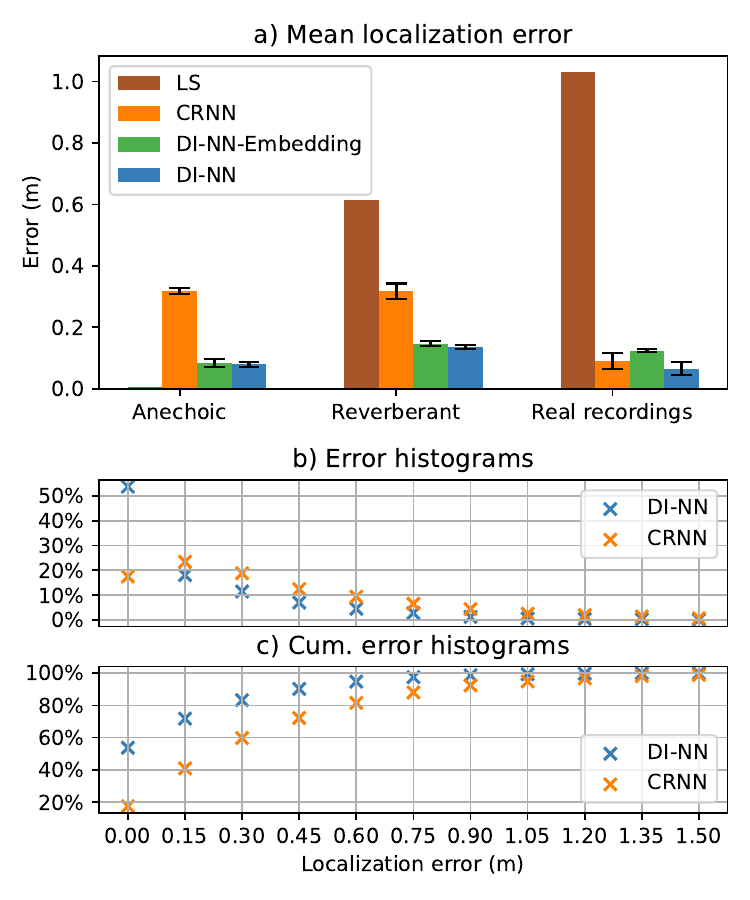}
    \end{center}
    \caption{(a) Mean localization error for DI-NNs and baselines on different datasets. (b) Normalized histogram comparison between the DI-NN and the CRNN baseline on the recorded dataset. (c) Cumulative version of (b).}
    \label{fig:errors}
\end{figure}

Figure~\ref{fig:errors}a compares the average error of our proposed \ac{DI-NN} and DI-NN-Embedding methods to the \ac{CRNN} and \ac{LS} baselines. To obtain statistically significant results, we train the \ac{DI-NN}, DI-NN-Embedding and \ac{CRNN} models four times independently for each experiment using random initial network parameters. The results shown in Fig. \ref{fig:errors} are averaged across the four times, with error bars showing the standard deviation across the runs. Conversely, as the \ac{LS} method is deterministic, it does not require multiple runs. 

A first remark is that although the \ac{LS} approach is very effective in the anechoic scenario, its performance is degraded on the other datasets, indicating its sensitivity to reverberation. The \ac{CRNN} outperforms the LS method in reverberant scenarios without knowledge of the microphone's coordinates. Interestingly, the \ac{CRNN} baseline is also obtains good localization performance on the recorded dataset, indicating that the network is able to infer the metadata to an extent when trained on a single room.   

However, by exploiting the microphone coordinates, the \ac{DI-NN} is shown to significantly improve the performance compared to the \ac{CRNN}. The most significant difference is observed in the anechoic case, where an improvement close to three times is obtained. In this case, the microphone coordinates are more useful as this information cannot be derived from the signals. In a reverberant room, however, the network might be able to use reflections to its advantage, as discussed in \cite{Ribeiro2010a}, to infer the microphone coordinates and making the metadata less useful.  Fig.~\ref{fig:errors}a also shows the errors obtained using the alternative DI-NN-Embedding architecture were similar to the \ac{DI-NN} in all scenarios, indicating no advantage in the proposed embedding, although it still allows the network to exploit the metadata.  

In turn, Fig.~\ref{fig:errors}b compares the normalized error histograms between our approach and the \ac{CRNN} baseline on the real recordings test dataset. The mode of the DI-NN's error is centred on the 0-15\,cm bin compared to the 15-30\,cm bin for CRNN's error. In other words, only the \ac{DI-NN} is median-unbiased. The cumulative distribution for the same data is shown in Fig.~\ref{fig:errors}c. While the DI-NN is shown to locate over 50\% and 80\% of the dataset samples with less than 15 and 45\,cm error, the \ac{CRNN} achieves the same errors for less than 20\% and 60\% of the data, respectively.

    \begin{table}[ht]
        \caption{Metadata sensitivity analysis} % title of Table
        \centering % used for centering table
        \begin{tabular}{c c c} % centered columns (4 columns)
        \hline\hline %inserts double horizontal lines
        Coord. std. (m) & Reverb. std. (ms) & Err. increase (\%) \\ [0.5ex] % inserts table
        %heading
        \hline % inserts single horizontal line
        0.01 & 0 & 0.05 \\ % inserting body of the table
        0.1 & 0 & 1.02 \\
        0.5 & 0 & 32.9 \\
        0 & 200 & 0.4 \\ [0.5ex] % [1ex] adds vertical space
        \hline %inserts single line
        \end{tabular}
        \label{table:sensitivity} % is used to refer this table in the text
    \end{table}
        
The results of the sensitivity study conducted in \autoref{sec:sensitivity} are displayed in Table~\ref{table:sensitivity}. The last column refers to the relative error increase between the perturbed case and the noiseless experiment conducted in \autoref{sec:simulated_reverb}. The results show that our approach is robust to the uncertainty inherent in practical measurements of the microphone coordinates and reverberation time estimates. The case where the microphone coordinates are disturbed by an extreme error of 0.5\,m (more than five times above typical errors) has been included to demonstrate the impact of including microphone coordinates for \ac{PSSL}, reiterating the importance and improved performance of metadata in our proposed fusion approach.

\begin{table}[ht]
    \caption{Metadata relevance analysis} % title of Table
    \centering % used for centering table
    \begin{tabular}{c c c c} % centered columns (4 columns)
    \hline\hline %inserts double horizontal lines
    Mic. coords. & Room dims. & RT60 & \% performance \\ [0.5ex] % inserts table
    %heading
    \hline % inserts single horizontal line
    \cmark & \cmark & \cmark & 100 \\ % inserting body of the table
    \cmark & \cmark & \xmark & 102 \\ % inserting body of the table
    \cmark & \xmark & \cmark & 100 \\ % inserting body of the table
    \xmark & \cmark & \cmark & 61 \\ % inserting body of the table
    \cmark & \xmark & \xmark & 104 \\ % inserting body of the table
    \xmark & \cmark & \xmark & 60 \\ % inserting body of the table
    \xmark & \xmark & \cmark & 47 \\ % inserting body of the table

    \hline %inserts single line
    \end{tabular}
    \label{table:relevance} % is used to refer this table in the text
\end{table}

Finally, the results of the metadata relevance analysis study described in \autoref{sec:relevance} are displayed in Table~\ref{table:relevance}. Each line represents a version of the DI-NN model trained on the reverberant dataset. The first three columns describe which metadata types are used in the model, and the last column shows the model performance relative to the model using all metadata, represented in the first line. The results show that the microphone coordinates are the most relevant for the model. In fact, using the microphone coordinates alone provides the best results. The results also indicate that the room dimensions are more relevant than the reverberation time in the absence of the microphone coordinates.

\subsection{Limitations and extensions} \label{sec:limitations}

Our approach exploits the metadata, such as the microphone coordinates and reverberation time and therefore this data must be known a priori or somehow measured. We have, however, shown that using this additional information is justified by a significant improvement in performance. While we have also assumed that the gains of the microphones are calibrated in our experiments, which may not be verifiable in practical scenarios, we have shown in \autoref{sec:real_recordings} that our model can perform well even when using uncalibrated microphones of the same kind. If calibration cannot be ensured, extracting gain invariant features from the signal pairs such as the cross spectra \cite{Xue2020} may be used as a preprocessing step.

We have also limited our scope to the localization of one static sound source using static microphones to focus on metadata fusion. However, extensions to moving sources and microphones could be possible by using smaller processing frames, for example. Another extension would be to estimate the three dimensional coordinates of the source. Finally, a possible extension for multiple source localization is expanding the output of DI-NN to a vector of size $2N$, where $N$ is the number of maximum sources, and performing Permutation Invariant Training (PIT) \cite{Dong2017}.  

\section{Conclusion}
In this work, we proposed \ac{DI-NN}, a simple yet effective way of jointly processing signals and relevant metadata using neural networks. Our results for the task of \ac{SSL} on multiple simulated and recorded scenarios indicate that the DI-NN is able to exploit successfully the metadata, as its inclusion reduced the mean localization error by a factor of at least two compared to the \ac{CRNN} baseline, as well as significantly improving localization results in comparison with the classical \ac{LS} algorithm in reverberant environments. Additional relevance and sensitivity studies revealed that the microphone coordinates the most important metadata, and that the DI-NN is robust to realistic noise in the metadata.

%% file: appendix.tex
\subsection{Validation of metadata independence between training and testing datasets} \label{sec:independence}

The datasets used in sections 4.1 and 4.2 are created entirely synthetically by generating random training, validation and testing samples. The attributes generated for each sample are the room's width and length, the coordinates of the four microphones, and the source coordinates. Additionally, in section 4.2, the room reverberation time is also randomly sampled. These values are then used to simulate the microphone recordings using the image source method. The only difference in the procedure for generating the training and testing sets is the random seed used for sampling values. Although highly unlikely, generating a test sample with the exact room dimensions, reverberation time, microphone and source coordinates as a sample in the training set could be possible and would violate the machine learning principle of having independent training and testing sets.

To assure the reader that this has not occurred in our experiments, we compute a distance metric $D$ between each testing sample and the entire training dataset. We focus on comparing the microphone coordinates between the training and testing sets and show that our approach has been validated against unseen metadata. Each sample comprises four microphone coordinates, each placed near one of the room's walls. We define the distance $d(i, j)$ between the $i$-th testing sample and $j$-th training sample as the sum of the distances of the microphone coordinates between the samples given by
\begin{equation} \label{eq:dist}
    d(i, j) = \sum_{k=1}^{4}\| \pmb{m}_i^k - \pmb{m}_j^k \|_2  \;,
\end{equation}
where $\pmb{m}_i^1$, $\pmb{m}_i^2$, $\pmb{m}_i^3$ and $\pmb{m}_i^4$ refer to the coordinates of the microphones located near the north, south, east and western walls of the room from the $i$-th sample and $\| \cdot \|$ denotes the $L_2$-norm.

\begin{figure}[h]
    \begin{center}
        \includegraphics[scale=0.465]{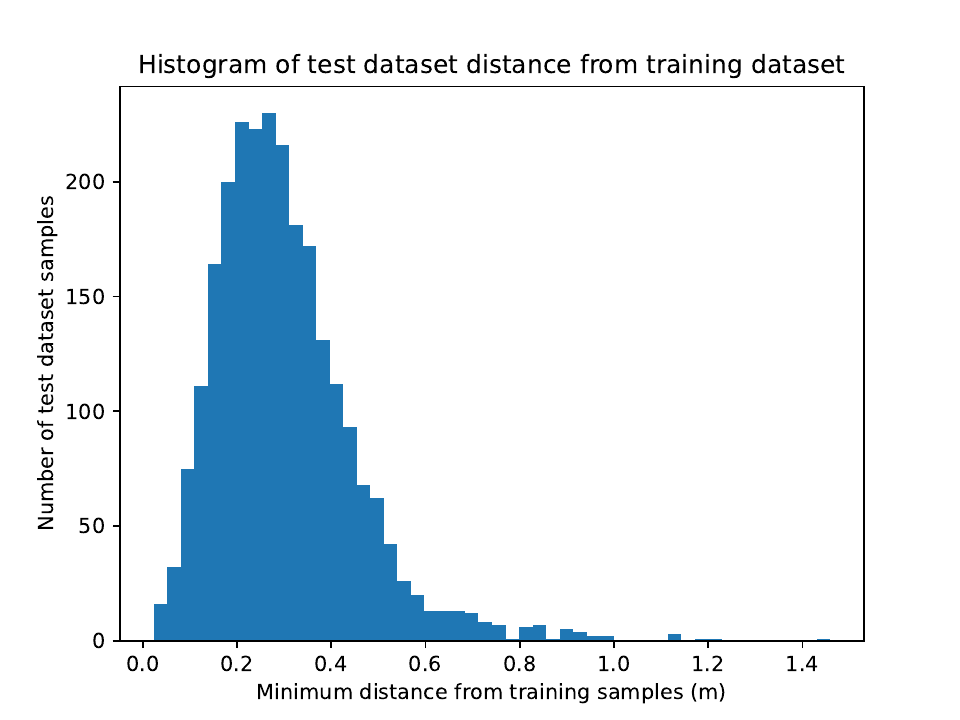} 
    \end{center}
    \caption{Distance between test dataset's microphone coordinates and training dataset.}
    \label{fig:hist}
\end{figure}

To measure the distance between the $i$-th testing sample and the entire training dataset, we compute \eqref{eq:dist} for every training sample $j$. We define the smallest distance $D(i)$ between the $i$-th testing sample and the entire training set as the minimum distance between $i$ and all training samples $j$, expressed as
\begin{equation} \label{eq:min_dist}
    D(i) = \min\limits_{j} \ \{d(i, j) \} \;.
\end{equation}
This measure quantifies the worst-case similarity between the $i$-th testing sample and the most similar sample in the entire training set. By plotting a histogram of $D(i)$ for every $i$-th sample in the testing set, we observe in Fig.~\ref{fig:hist} that no training microphone configuration appeared in the testing set. Moreover, the average minimum distance between the testing and training sets is around 30\,cm. Besides having different microphone coordinates, we like to emphasize that the room dimensions and reverberation time also vary from sample to sample, increasing the differences between training and testing sets even further.

